# Magnetotransport of SrIrO$_3$ films on (110) DyScO$_3$

A. K. Jaiswal[1,2], A. G. Zaitsev[1], R. Singh[2], R. Schneider[1], and D. Fuchs[1,*]

[1] Institute for Solid-State Physics, Karlsruhe Institute of Technology, 76021 Karlsruhe, Germany
[2] Department of Physics, Indian Institute of Technology Delhi, 110016 New Delhi, India

Epitaxial perovskite (110) oriented SrIrO$_3$ (SIO) thin films were grown by pulsed laser deposition on (110) oriented DyScO$_3$ (DSO) substrates with various film thickness $t$ (2 nm < $t$ < 50 nm). All the films were produced with stoichiometric composition, orthorhombic phase, and with high crystallinity. The nearly perfect in-plane lattice matching of DSO with respect to SIO and same symmetry result in a full epitaxial in-plane alignment, i.e., the $c$-axis of DSO and SIO are parallel to each other with only slightly enlarged $d_{110}$ out-of-plane lattice spacing (+0.38%) due to the small in-plane compressive strain caused by the DSO substrate. Measurements of the magnetoresistance *MR* were carried out for current flow along the [001] and [1-10] direction of SIO and magnetic field perpendicular to the film plane. *MR* appears to be distinctly different for both directions. The anisotropy $MR_{001}/MR_{1-10} > 1$ increases with decreasing $T$ and is especially pronounced for the thinnest films, which likewise display a hysteretic field dependence below $T^* \approx 3$ K. The coercive field $H_c$ amounts to 2-5 T. Both, $T^*$ and $H_c$ are very similar to the magnetic ordering temperature and coercivity of DSO which strongly suggests substrate-induced mechanism as a reason for the anisotropic magnetotransport in the SIO films.

## I. INTRODUCTION

Quantum materials with competing strength of various interactions have gained increasing interests in recent years due to their rich phase diagram and possible application of realizing new quantum states [1,2]. Iridium based 5d- transition metal oxides (TMOs) display large spin-orbit coupling (SOC) compared to 3d- & 4d- TMOs due to the heavy Ir with atomic number $Z$ = 77 (SOC $\propto Z^4$) [3]. The delicate interplay of spin-orbit coupling, electron correlations, and crystal field energy makes iridates a prototype system to explore exotic phases originating from the competition of these interactions [4-14].

*Corresponding author: dirk.fuchs@kit.edu

Dimensionality controlled physical properties have been observed in the Ruddlesden-Popper (RP) family of strontium iridates ($Sr_{n+1}Ir_nO_{3n+1} \equiv SrO(SrIrO_3)_n$) where $n$ layers of $SrIrO_3$ are intercalated between SrO layers along the $c$-axis direction [15]. The end member $SrIrO_3$ ($n = \infty$) has a semimetallic paramagnetic ground-state. An increase of $n$ results in an increase of hybridization of Ir $5d$ and O $2p$ orbitals which consequently pushes $SrIrO_3$ (SIO) to a semimetallic state [16]. Thus, SIO is very close to a metal-to-insulator transition and magnetic order. Therefore, the properties of SIO are expected to be susceptible to external perturbations.

At ambient pressure, the monoclinic phase $C2/2$ (15) of SIO is energetically favorable while orthorhombic perovskite phase $Pbnm$ (62) requires synthesis under high pressure ($p \approx 40$ kbar). However, it is possible to stabilize the orthorhombic phase at ambient pressure using thin-film growth technology. Recently, people have successfully grown thin films of SIO on various lattice-matched substrates [17-19]. Growth conditions, underlying substrate, and film thickness have a profound effect on the electronic properties of SIO films [16,20]. A metal-to-insulator transition (MIT) has been observed for SIO films with thickness $t \leq 3$-unit cells on $SrTiO_3$ (001) where suppression of in-plane octahedral rotation opens a charge gap at the Fermi level [21]. In literature, there are few reports of magnetic ordering in SIO based interfaces and superlattices [22,23]. For [$(SrIrO_3)_m$, $SrTiO_3$] ($m = 1,2,3$ and $\infty$) superlattices, where $m$ is the number of SIO unit cells, a semimetal to insulator and magnetic phase transition has been observed simultaneously, for $m \approx 3$, indicating the correlation between charge gap and magnetic ordering [22].

In this study, we report on the magnetotransport of SIO films on DSO (110) substrates for various film thickness $t$. Magnetoresistance $MR$ is positive and shows sharp cusp at lower fields hinting to SOC-induced weak-antilocalization. MR displays different behavior along the [001] and [1-10] direction. In contrast to films with $t \geq 9$ nm, SIO film with $t = 2.3$ nm show a hysteretic behavior below $T^* \approx 3$ K. The measurements hint to a substrate-induced magnetic order in the SIO film.

II. EXPERIMENTAL

(110) oriented perovskite SIO films were grown on $DyScO_3$ (110) substrates (5×5 mm$^2$ from CrysTec GmbH) with various film thickness $t$ (2nm < $t$ < 50 nm) using a pulsed laser deposition system with a KrF excimer laser ($\lambda = 248$ nm). Details on film growth are described elsewhere [24]. A capping layer of STO (~ 4 nm) was deposited *in-situ* to protect and prevent the SIO film surfaces from any possible degradation. Structural properties of the films such as film thicknesses $t$, roughness $r$, crystallinity and out of plane lattice spacing $d_{110}$ were analyzed by x-ray reflectivity (XRR) and diffraction (XRD) measurements using a four-circle diffractometer (Bruker D8 Discover). Stoichiometric film composition was confirmed by Rutherford backscattering spectrometry.

To study electronic transport, microbridges in Hall bar geometry were patterned along the two orthogonal [1-10] and [001] crystallographic in-plane directions by standard UV-photolithography and argon-ion milling. The dimensions of the Hall bars are $100 \times 20$ µm$^2$. Patterned samples were post-annealed in flowing $O_2$ for 5 h to avoid possible oxygen deficiency and parasitic conductivity induced by argon-ion milling. Electrical contacts for the transport measurement were produced by ultrasonic Al-wire bonding. Resistance measurements were carried out in standard four-point probe (FPP) geometry using a Quantum Design physical property measurement system (PPMS) equipped with a 14-T superconducting magnet. Magnetization measurements were performed by means of a Quantum Design superconducting quantum interference device (SQUID).

III. RESULTS AND DISCUSSION

In Fig. 1, the x-ray reflectivity and θ/2θ x-ray diffraction in the vicinity of the (110) reflection of DSO are presented. From the fitting of measured XRR curves the critical angle ($α_c$), film thickness ($t$) and surface roughness (r) are deduced which reveal a smooth layer-by-layer growth with negligible surface roughness. Symmetric XRD scans of SIO show a single crystalline phase. With decreasing $t$, the peak maximum of the (110) film reflection is shifted toward lower 2θ values, i. e., larger d-spacing due to the in-plane compressive strain caused by the DSO substrate. The out-of-plane lattice spacing ($d_{110}$) increases from 3.96 Å for $t$ = 50 nm to 4.09 Å for $t$ = 2.3 nm. Laue oscillations on both sides of the (110) reflection evidence a layered growth.

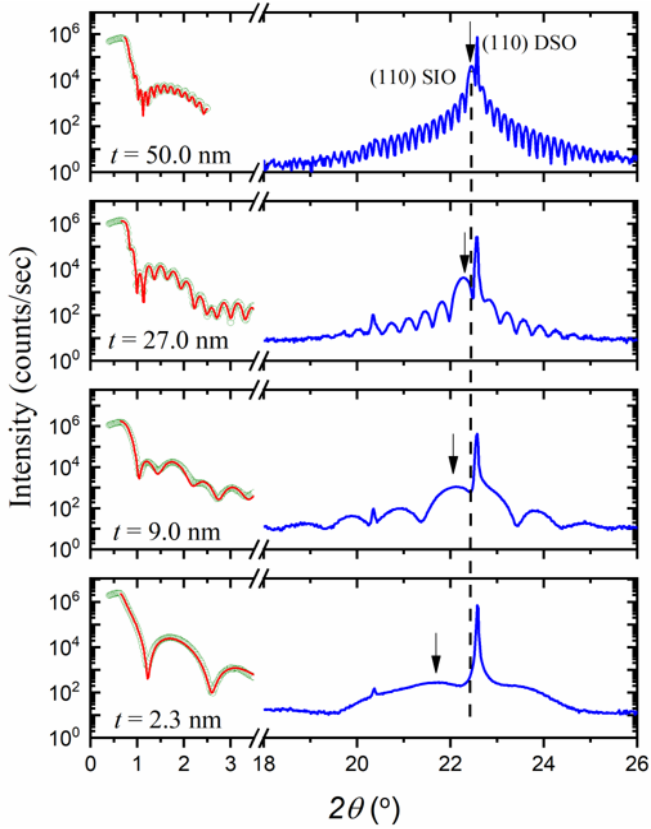

FIG.1. X-ray reflectivity (green symbols) and x-ray diffraction (blue lines) for the SIO films deposited on DSO (110) for different film thickness $t$. Red solid lines are fits to the reflectivity. Film peak position of the (110) reflection is indicated by arrow. The dashed line indicates the $2\theta$ peak position of the (110) reflection of bulk SIO.

Fig. 2 shows the normalized resistivity $\rho_{001}(T)/\rho_{001}(300K)$ and $\rho_{1\text{-}10}(T)/\rho_{1\text{-}10}(300K)$ of SIO films for the two orthogonal in-plane directions for various film thickness $t$. Generally, $\rho_{001}(T)$ is distinctly smaller compared to $\rho_{1\text{-}10}(T)$. In Fig. 2 (d), we have plotted the ratio of the normalized resistivity, $r_n = [\rho_{1\text{-}10}(T)/\rho_{1\text{-}10}(300\text{ K})]/[\rho_{001}(T)/\rho_{001}(300\text{ K})]$, as a function of $T$. The anisotropic behavior can be discussed in terms of the $T$-dependence of the structural in-plane anisotropy, $\sqrt{(a(T)^2 + b(T)^2)}/c(T)$, which for bulk SIO [24, 25] looks very similar to the $T$-dependence of $r_n$. Interestingly, $r_n(T)$ does not change significantly with decreasing $t$ for $t > 2.3$ nm. However, for $t = 2.3$ nm, $r_n(T)$ drops down to about 1.08 at 4 K, indicating less anisotropic transport. Possible reason for that is very likely an increase of epitaxial strain to the SIO layer, which is naturally expected with decreasing film thickness. Mismatch of thermal expansion between SIO and DSO may also lead increasingly to $T$-dependent structural changes, i. e., orthorhombic distortion, with decreasing $t$, which may affect resistivity anisotropy in SIO considerably. The measurements demonstrate the impact of the substrate material on the electronic transport and anisotropy of thin ($t \leq 2.3$ nm) SIO films.

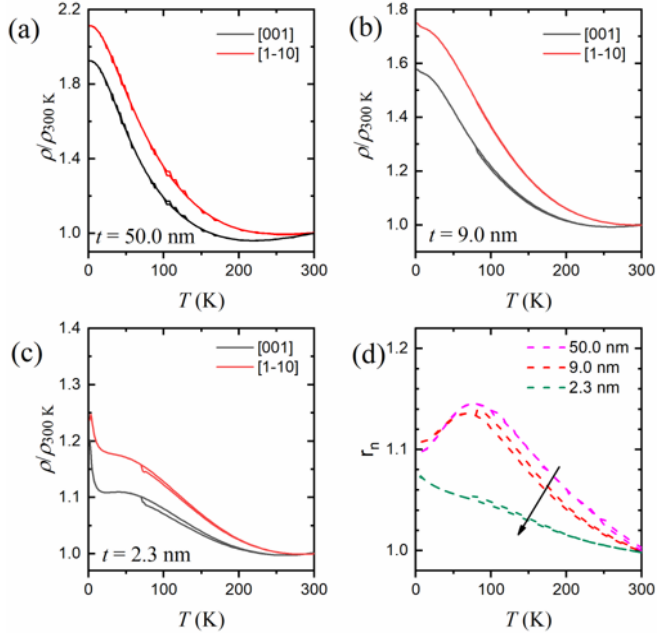

FIG. 2. Normalized resistivity $\rho(T)/\rho(300$ K$)$ as a function of $T$ for the [001] and [1-10] directions and $t = 50$ nm (a), 9 nm (b), and 2.3 nm (c). The ratio of the normalized resistivity, $r_n = [\rho_{1-10}(T)/\rho_{1-10}(300$ K$)]/[\rho_{001}(T)/\rho_{001}(300$ K$)]$, as a function of $T$ for different SIO film thickness $t = 50$, 9, and 2.3 nm (d). The arrow indicates the reduction of $r_n$ with decreasing $t$, most prominent for $t = 2.3$ nm.

In the following, we will focus on the magnetotransport of SIO. Magnetoresistance measurements were performed with magnetic field $B$ perpendicular to the film surface. Figure 3 shows magnetoresistance $MR = \frac{R(B)-R(0)}{R(0)}$ of of patterned SIO microbridges for various film thickness $t$. $MR$ is always positive and displays an anisotropic behavior along the [1-10] and [001] direction with $MR_{001}/MR_{1-10} > 1$. Generally, $MR$ increases with decreasing $T$ and increasing $B$. For the thinnest film ($t = 2.3$ nm), the anisotropy ($MR_{001}/MR_{1-10}$) shows a distinct increase. In addition, a field-dependent butterfly-shaped hysteresis occurs for different field-sweep direction at $T = 2$ K. Interestingly, the field-dependence changes significantly if $T$ is increased to 5 K, see Fig. 3 (d). To document the hysteresis-behavior more clearly, we have plotted again $MR$ at $T = 2$ K for the low field range -2.5 T $< B <$ 2.5 T in Fig. 4.

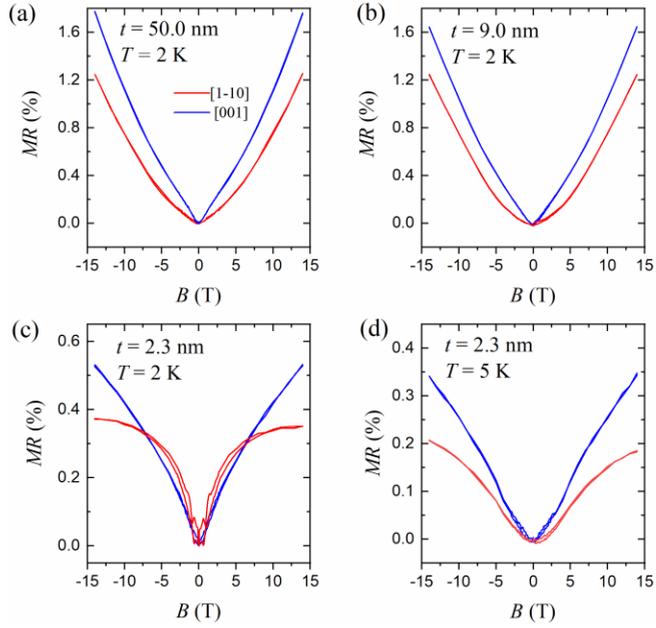

FIG. 3. Magnetoresistance $MR$ at $T = 2$ K for the [1-10]- (red line) and [001]-direction (blue line) of a patterned SIO film on DSO with $t = 50.0$ nm (a), $t = 9.0$ nm (b), and $t = 2.3$ nm (c). (d) $MR$ for $t = 2.3$ nm at $T = 5$ K.

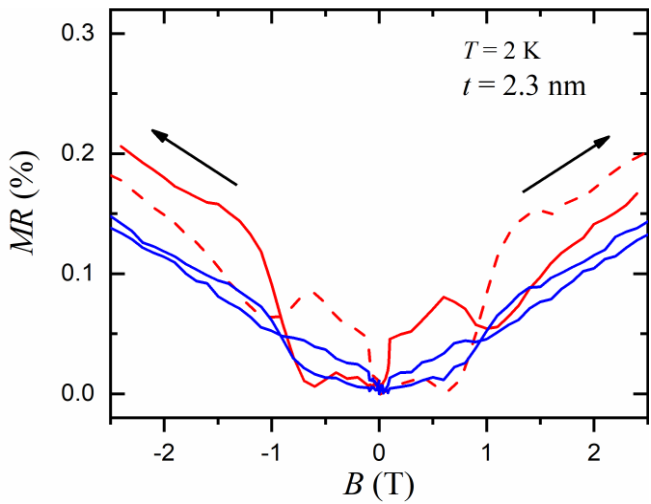

FIG. 4. Low-field MR for a thin SIO film ($t = 2.3$ nm) on DSO at $T = 2$ K for the [1-10] (red color) and the [001] (blue color) direction. The arrows indicate upward (dashed lines) and downward (solid lines) sweep-direction for the [1-10] direction.

Magnetic field sweep direction is indicated in the figure by arrows. The hysteresis is obviously more pronounced for the [1-10] direction. The minima around 0.5 T (for upward sweep-direction) and -0.5 T (for downward sweep-direction) are anomalies with pronounced deviations from classical $B^2$-dependence of *MR* which are very likely caused by weak antilocalization. Such features, typically occurring at $B = 0$ for SIO on STO are well known for thin SIO films [20] and are caused by the strong SOC of the iridates. Interestingly, the hysteretic behavior of *MR* disappears for $T ≥ 5$ K which hints to magnetic order in the film below 5 K. Such an anomalous and anisotropic behavior of *MR* is not observed for SIO films on STO, which strongly suggests influence of the DSO substrate material. DSO displays strong paramagnetic behavior with antiferromagnetic ordering at $T_N = 3.1$ K [26,27]. In addition, the Dy-moments display canted alignment, which gives rise to a net magnetic moment and hence hysteretic behavior below $T_N$ [26]. In addition, DSO displays strong magnetic anisotropy with [1-10] direction as easy axis and [001] direction being the hard axis.

The simultaneous appearance of hysteretic effects below 5 K on resistivity and magnetic moment for SIO and DSO, respectively, strongly hints to a substrate-induced mechanism as a reason for the anisotropic magnetotransport in SIO. The strong magnetic anisotropy of DSO likely results in a large spin-polarization of SIO charge carriers along the [1-10] direction, i. e., the easy axis direction of DSO, if magnetic field is applied. Consequently, spin-flip scattering becomes less effective along the [1-10] direction which may explain the reduced *MR* compared to that of the [001] direction. Reducing the film thickness makes substrate induced effects more transparent, and thus, anisotropic MR is stronger in the thinnest film. Possibly, substrate induced spin-polarization in SIO is significant in the vicinity of the substrate/film interface and hence limited to the first few SIO monolayers. More detailed information on the thickness of the spin-polarized layer has to be figured out by further experiments.

## IV. SUMMARY AND CONCLUSIONS

The magnetotransport properties of SIO films on DSO substrates have been investigated. The measurements demonstrate that the *MR* is always positive and displays weak-antilocalization for $T < 10$ K and $B < 2$ T. The *MR* shows anisotropic behavior for the [1-10] and [001] direction with $MR_{1-10} < MR_{001}$. In addition, a butterfly-shaped hysteresis is observed for the thinnest SIO film ($t = 2.3$ nm) on DSO below 5 K. The observed anisotropy and hysteretic behavior of the *MR* evidence substrate-induced impact on the magnetotransport in thin SIO films.

ACKNOWLEDGMENTS

We would like to thank R. Thelen and the Karlsruhe Nano Micro Facility (KNMF) for providing us the Atomic Force Microscopy facility. We also thank J. Schubert from the Peter Grünberg Institut, Forschungzentrum Jülich, for Rutherford backscattering spectrometry measurements. AKJ acknowledges DAAD for financial funding within India IIT Master Sandwich Programme (57434206). AKJ and DF also acknowledge S. Mukherjee and K. Sen for fruitful discussions.

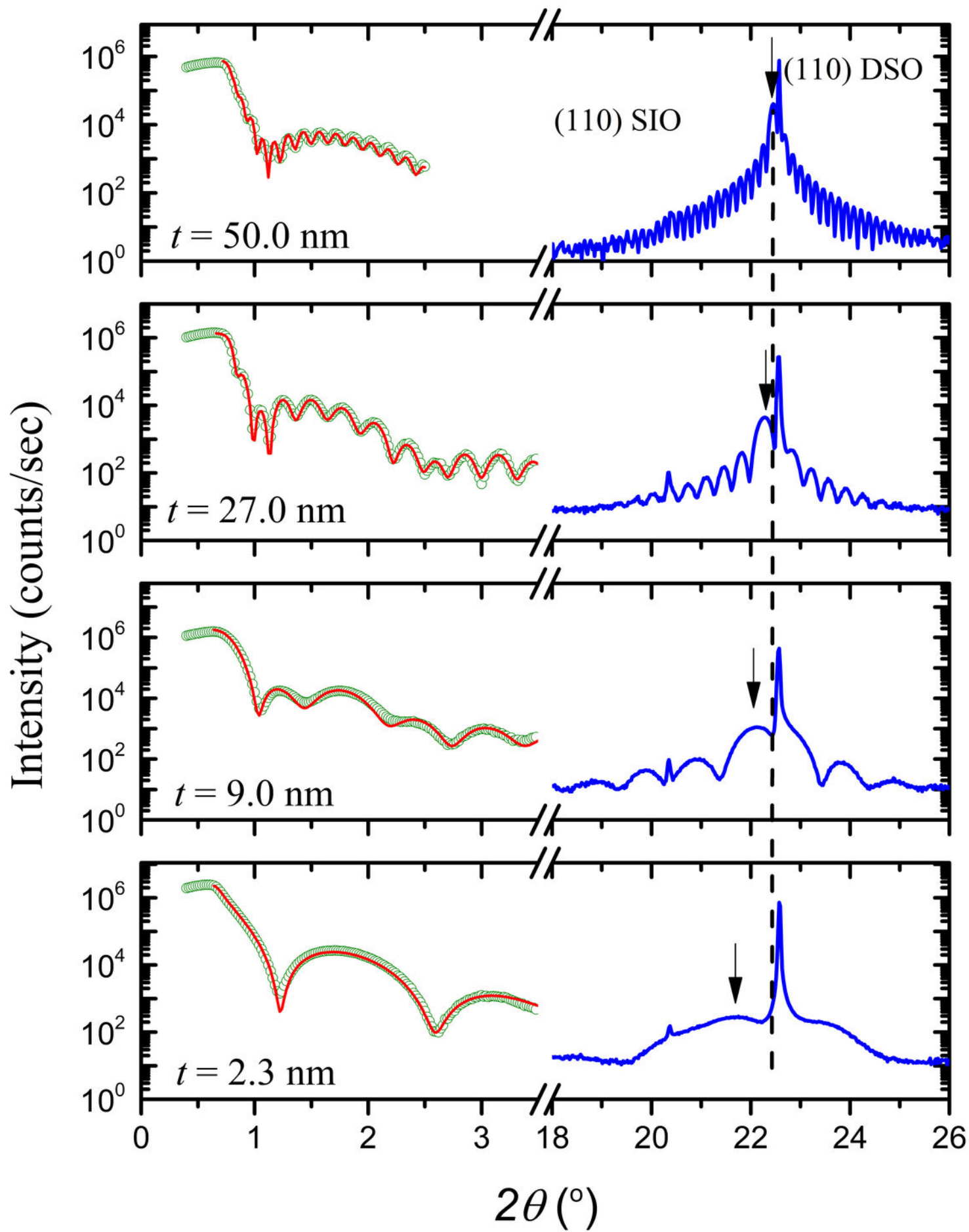

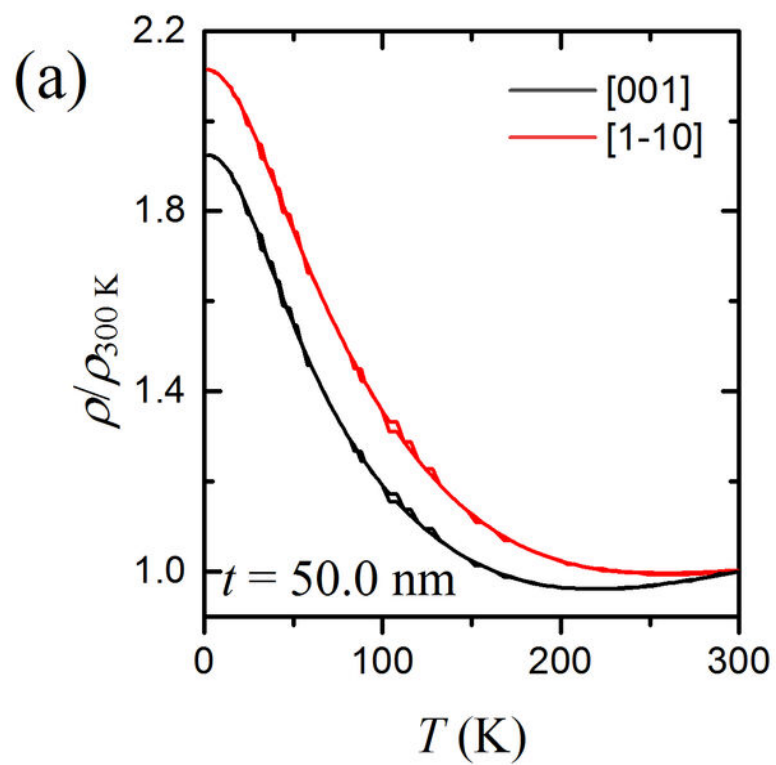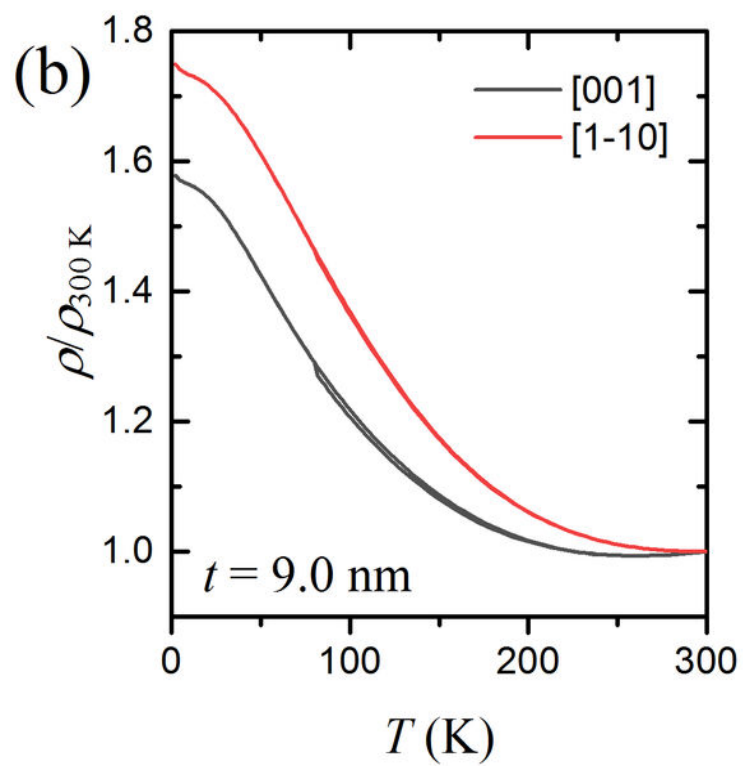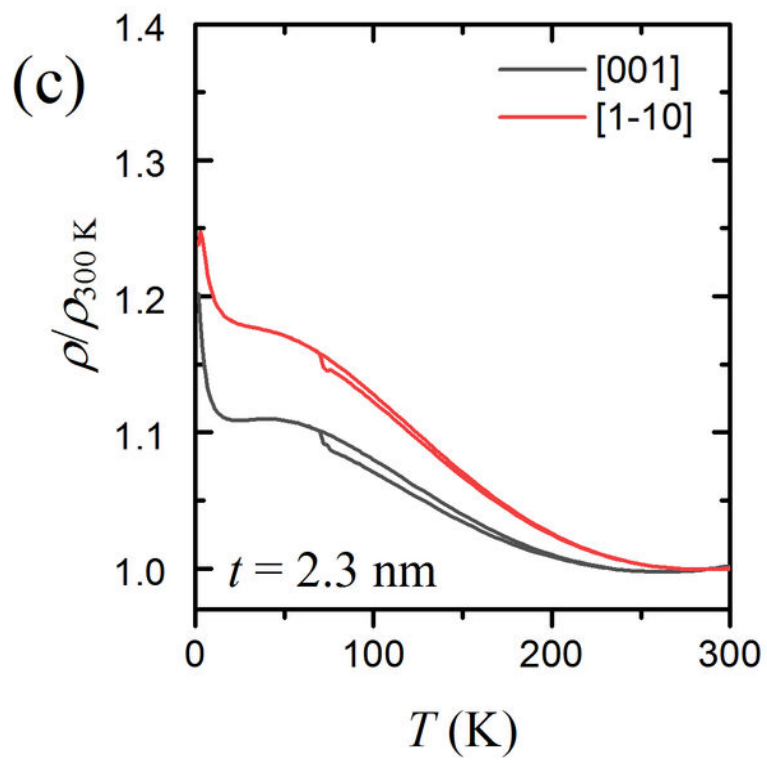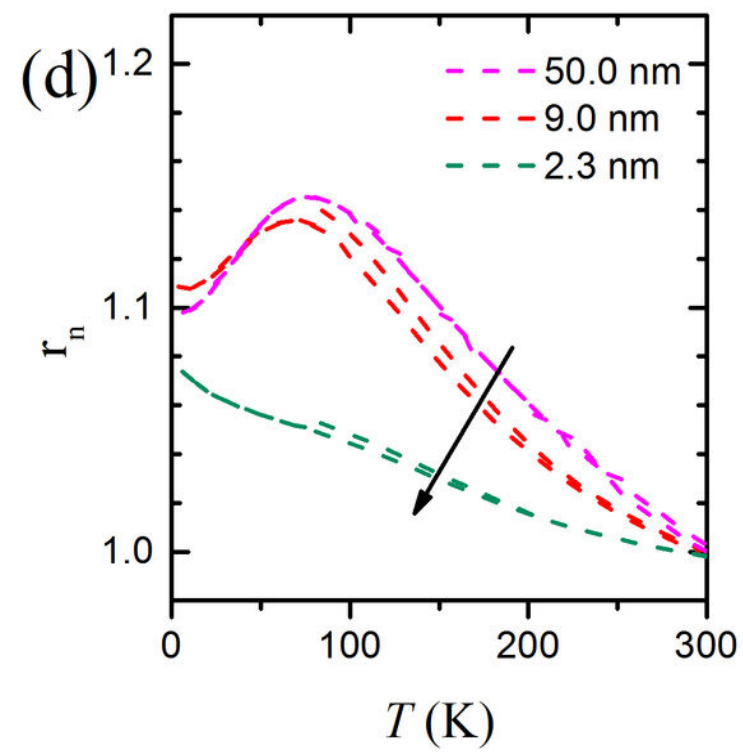

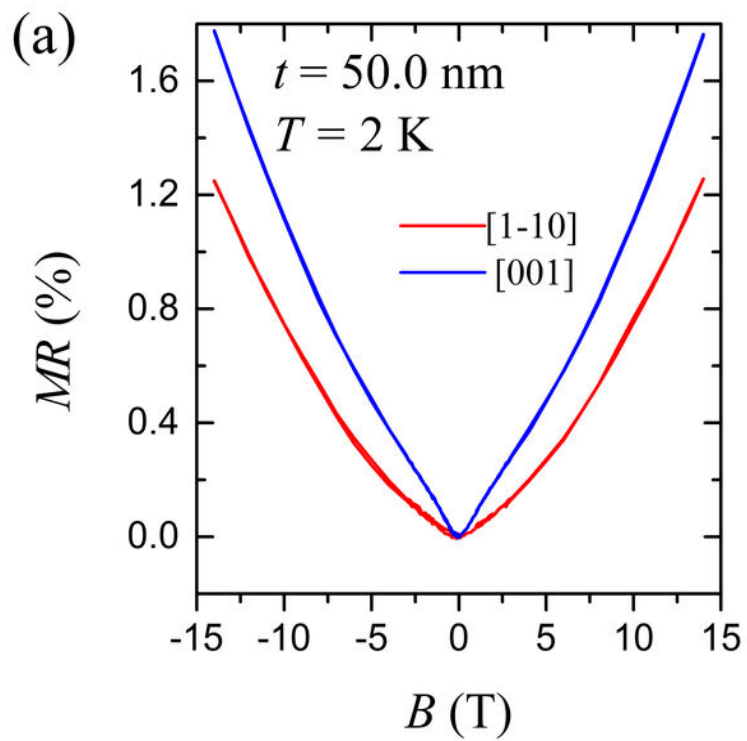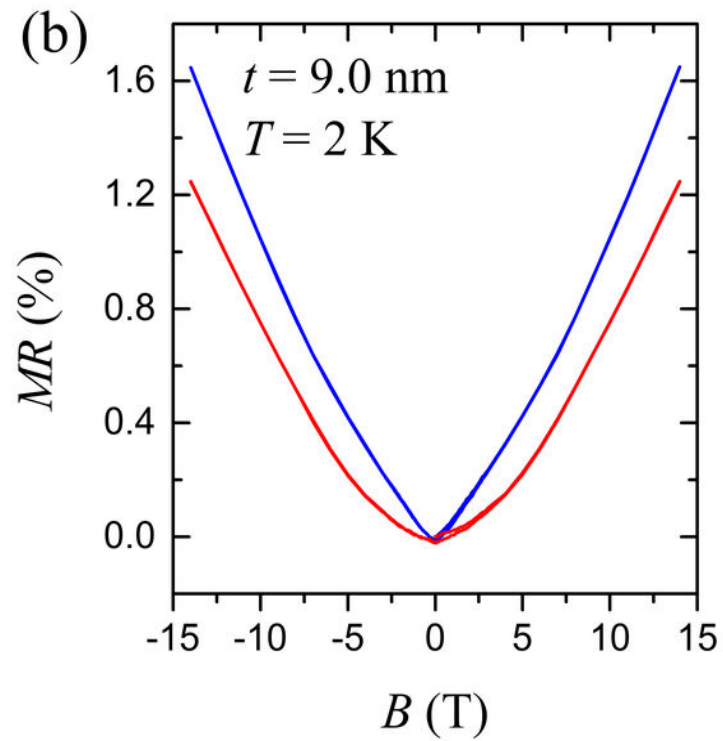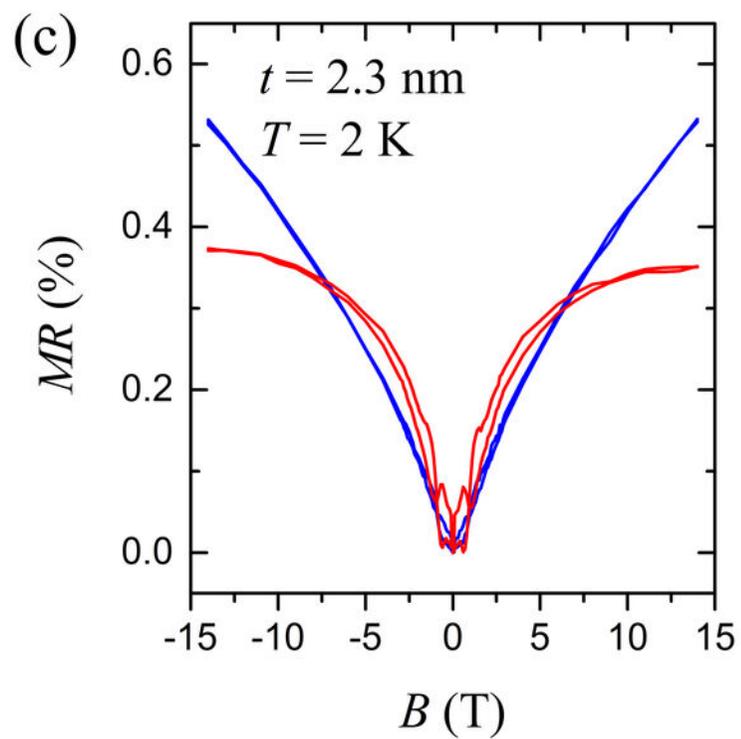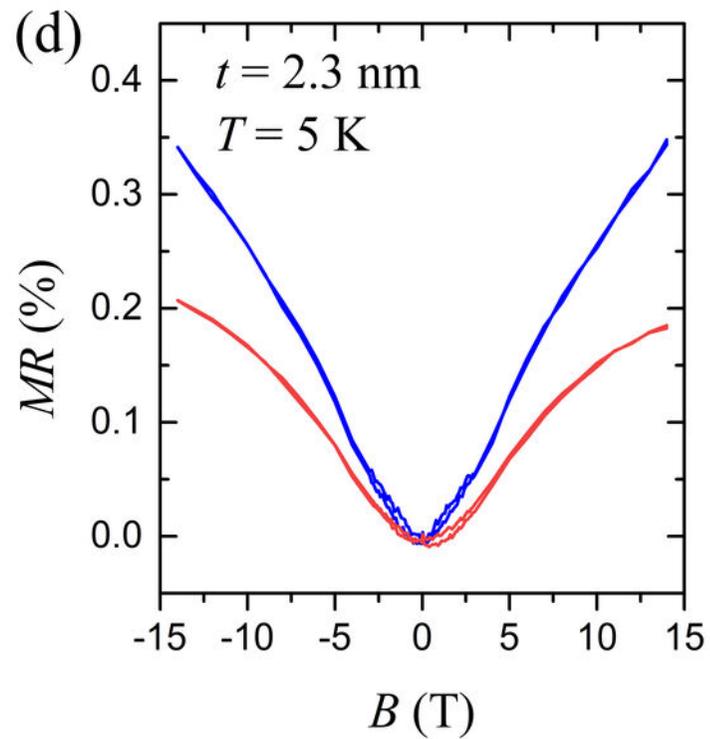

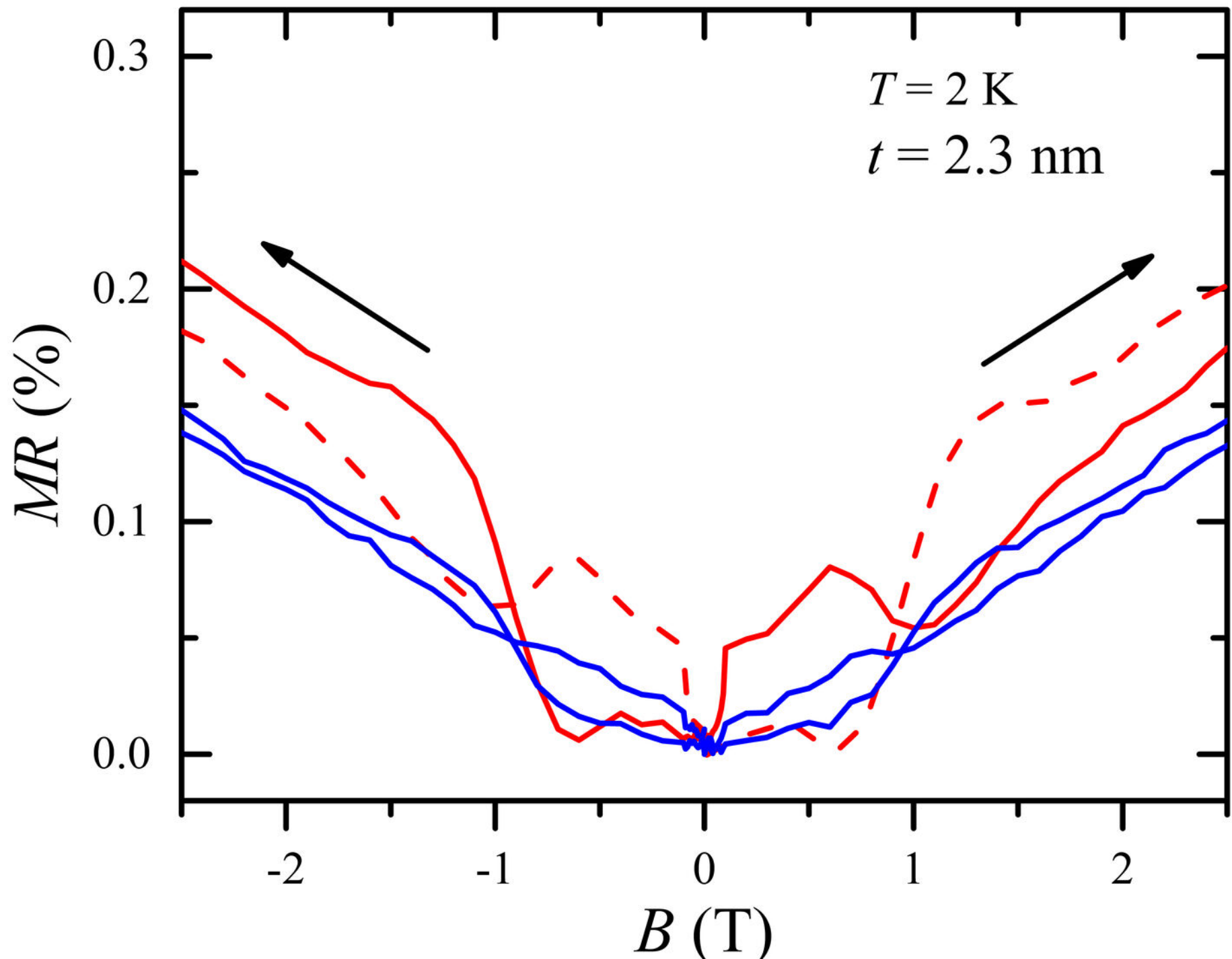